\PassOptionsToPackage{table,xcdraw}{xcolor}

\documentclass[sigconf]{acmart}

\copyrightyear{2024}
\acmYear{2024}
\setcopyright{acmlicensed}\acmConference[ASSETS '24]{The 26th International
ACM SIGACCESS Conference on Computers and Accessibility}{October 27--30,
2024}{St. John's, NL, Canada}
\acmBooktitle{The 26th International ACM SIGACCESS Conference on
Computers and Accessibility (ASSETS '24), October 27--30, 2024, St. John's,
NL, Canada}
\acmDOI{10.1145/3663548.3675663}
\acmISBN{979-8-4007-0677-6/24/10}

\acmConference[ASSETS '24]{Make sure to enter the correct
  conference title from your rights confirmation email}{October 27--30, 2024}{St. John's,
NL, Canada}
\usepackage{subfigure}

\usepackage{graphicx}
\usepackage{multirow}
\usepackage{pifont}
\usepackage{lscape}
\usepackage{tabularray}

\setcopyright{none}
\begin{document}

\title[Accessible Nonverbal Cues to Support Conversations in VR for Blind and Low Vision People]{Accessible Nonverbal Cues to Support Conversations in VR for Blind and Low Vision People}

\author{Crescentia Jung}
\authornote{Crescentia Jung and Jazmin Collins are co-authors and contributed equally to this research.}
\email{cj382@cornell.edu}
\affiliation{%
  \institution{Cornell University}
  \streetaddress{2 W Loop Rd}
  \city{New York}
  \state{NY}
  \country{USA}
  \postcode{10044}
}
\author{Jazmin Collins}
\authornotemark[1]
\email{jc2884@cornell.edu}
\affiliation{%
  \institution{Cornell University}
  \streetaddress{2 W Loop Rd}
  \city{New York}
  \state{NY}
  \country{USA}
  \postcode{10044}
}
\author{Ricardo E. Gonzalez Penuela}
\email{reg258@cornell.edu }
\affiliation{%
  \institution{Cornell University}
  \streetaddress{2 W Loop Rd}
  \city{New York}
  \state{NY}
  \country{USA}
  \postcode{10044}
}
\author{Jonathan Isaac
Segal}
\email{jis62@cornell.edu}
\affiliation{%
  \institution{Cornell University}
  \streetaddress{}
  \city{Ithaca}
  \state{NY}
  \country{USA}
  \postcode{14850}
}
\author{Andrea Stevenson Won}
\email{asw248@cornell.edu}
\affiliation{%
  \institution{Cornell University}
  \streetaddress{}
  \city{Ithaca}
  \state{NY}
  \country{USA}
  \postcode{14850}
}
\author{Shiri Azenkot}
\email{shiri.azenkot@cornell.edu}
\affiliation{%
  \institution{Cornell Tech}
  \streetaddress{2 W Loop Rd}
  \city{New York}
  \state{NY}
  \country{USA}
  \postcode{10044}
}

\renewcommand{\shortauthors}{Jung and Collins, et al.}

\begin{abstract}
Social VR has increased in popularity due to its affordances for rich, embodied, and nonverbal communication. However, nonverbal communication remains inaccessible for blind and low vision people in social VR. We designed accessible cues with audio and haptics to represent three nonverbal behaviors: eye contact, head shaking, and head nodding. We evaluated these cues in real-time conversation tasks where 16 blind and low vision participants conversed with two other users in VR. We found that the cues were effective in supporting conversations in VR. Participants had statistically significantly higher scores for accuracy and confidence in detecting attention during conversations with the cues than without. We also found that participants had a range of preferences and uses for the cues, such as learning social norms. We present design implications for handling additional cues in the future, such as the challenges of incorporating AI. Through this work, we take a step towards making interpersonal embodied interactions in VR fully accessible for blind and low vision people. 
\end{abstract}





\maketitle
\section{Introduction}
Social virtual reality (VR) is growing in popularity but remains inaccessible to blind and low vision (BLV) people. Social VR has thousands of daily users that participate in a variety of social activities from informal parties to remote business meetings. In addition, the embodied nature of social VR allows people to communicate with nonverbal behaviors such as gestures and facial expressions. However, these behaviors are generally rendered visually, and this limits BLV people’s access to social information, excluding them from full participation.

Researchers have explored how to make social VR more accessible for BLV people \cite{zhang2022s, collins2023guide, gualano2023invisible, zhao2018enabling, ji2022vrbubble}. For example, Zhang et al. explored avatar diversity and the importance of self-presentation of people with disabilities in social VR \cite{zhang2022s}. Additionally, Collins and Jung et al. studied the use of a sighted guide as an accessibility tool in social VR \cite{collins2023guide}. 

However, there has been little exploration of making nonverbal behaviors accessible in social VR. Some have investigated what kinds of behaviors BLV people wish to be made accessible in VR, as well as initial guidelines for the best ways of making certain behaviors accessible. For example, Wieland et al. \cite{wieland2022non} conducted a user study to identify which nonverbal behaviors BLV people used in conversations with sighted partners and should be carried over into social VR, such as facial expressions, head movements, and gaze. Similarly, Collins and Jung et al. \cite{collins2023making} worked with a BLV person to identify best practices for representing gaze in VR. Ji et al.’s \cite{ji2022vrbubble} “VRBubble” used audio beacons to indicate proxemic information about avatars to improve BLV people’s awareness of others. 

Prior works such as those by Collins and Jung et al. \cite{collins2023making} and Ji et al. \cite{ji2022vrbubble} have designed accessibility features for one type of nonverbal behavior at a time. The participants evaluating these systems also experienced only limited, pre-recorded conversations with agent-avatars, instead of actual people. This is understandable, considering the complexity and the immense range of the nonverbal behaviors real people use in conversation. Trying to design representations of multiple behaviors at a time is challenging, and may involve various problems such as determining which behaviors to represent, the best ways to represent them, and ensuring all representations can be used at once without becoming overwhelming. However, only having one accessible behavior during pre-recorded interactions does not represent real conversations, or provide enough nonverbal information to fully support conversation. BLV users need access to multiple nonverbal behaviors to support conversations with others in VR.

To address this need, we sought to design multiple accessible nonverbal cues that could be used simultaneously in social VR for BLV people. Specifically, we aim to address the following research question: \textbf{What set of accessible nonverbal cues can support BLV users in conversations with others in VR?} We use the word “cue” here and in the remainder of our paper to refer to a representation of a nonverbal behavior, while the word “behavior” refers to the nonverbal behavior itself. 

We conducted a user-centered design process to design accessible audio and haptic cue representations of nonverbal behaviors. We iterated on five cue designs with six BLV participants. At the end of the process, we found that five cues were overwhelming for participants, and narrowed our focus to three final cues: eye contact, head nodding, and head shaking.

We then evaluated this set of three cues with 16 participants in social VR conversation tasks. Each participant completed two conversation tasks: one with cues and one without. We found that our nonverbal cues significantly improved participants’ accuracy in detecting certain social behaviors about their conversation partners without distracting them from the conversation. Participants also reported feeling more confident about “reading the room”, especially knowing how much attention they were receiving. Our discussions with participants yielded design implications for how nonverbal behaviors might be represented accessibly in the future, including more behaviors participants wanted to access, and potential ways of managing larger sets of accessible cues for these behaviors. 

Our contributions include the design and evaluation of three accessible nonverbal cues in VR and novel design implications for sets of accessible nonverbal cues.
\section{Related Work}
\subsection{Accessible VR for BLV People}
VR experiences tend to be over-reliant on visual information, making them inaccessible for BLV people. To address this, researchers have explored how to make virtual environments and the content within them l perceivable to BLV users. Much of this work has focused on providing \textit{environmental} information via haptics or audio to enhance BLV users’ individual experiences in VR, but far less has focused on providing \textit{social} information to enhance social experiences with other people.

Many researchers have investigated how to provide information about a VR environment’s layout to support BLV people’s navigation \cite{guerreiro2023design, andrade2018echo, hao2022detect, nair2021navstick, siu2020virtual, zhang2020exploring, zhao2018enabling, gonccalves2023inclusive}. For instance, Andrade et al. \cite{andrade2018echo} designed EchoHouse, a virtual environment where users navigate with “echolocation”, using real-time audio cues emitted from objects in the space. More recently, Collins and Jung et al. \cite{collins2023guide} proposed supporting navigation through sighted guides. In their work, they implemented a system that allowed a BLV user to pair up with a sighted human guide to do visual interpretation of and navigate any virtual space together. 

Besides navigation, researchers have explored ways of conveying details about objects in virtual environments through audio and haptics \cite{kornbrot2007roughness, nikolakis2004cybergrasp, sinclair2019capstancrunch, zhao2019seeingvr, gonzalez2022hands}. Zhao et al. \cite{zhao2019seeingvr} created a developer toolkit called SeeingVR, which offers 14 different visibility enhancements to make VR experiences accessible for users with low vision (e.g., magnification, highlighting salient objects, font size adjustment, and others). Researchers have also developed approaches using haptics and gestures for blind users, such as Penuela et al.’s \cite{gonzalez2022hands} haptic gloves allow BLV users to perceive the shape of objects through force feedback and elicit descriptions about objects in the environment through gestures.

Some researchers have also created accessible VR games that use nonvisual modalities to help BLV users complete complex objectives in virtual environments \cite{bailenson2004transformed, gluck2021racing, gluck2020implementing, lumbreras1999interactive, morelli2010vi, wedoff2019virtual, nair2022spatial}. For example, Wedoff et al. \cite{wedoff2019virtual} designed Virtual Showdown, a VR audio game that provides spatialized audio cues to help players find, move towards, and hit a ball back to an opponent.

While the above efforts have taken an important step towards making \textit{individual} VR experiences accessible for BLV users, they do not account for the social aspects of environments where multiple users are present, like social VR applications. These applications are often designed so users interact with each other to complete shared objectives or bond with each other in conversations. In such cases, communicating social information about other people is more important  to support BLV users’ abilities to participate in the virtual space. Without access to social information about other people, BLV users may remain isolated in social VR spaces, even with accessibility enhancements for individual experiences. Our work seeks to address this gap by supporting BLV people’s access to a key aspect of communication in social VR: nonverbal behaviors.
\subsection{Nonverbal Behaviors in Social VR}
Nonverbal behaviors are an important aspect of social VR and have been explored extensively in prior work. However, this work has largely focused on understanding how they are used, rather than how to make them accessible.

Researchers have explored which nonverbal behaviors are most commonly used in social VR \cite{maloney2020talking, tanenbaum2020make}. Maloney et al. \cite{maloney2020talking} observed people’s use of nonverbal behaviors on social VR applications and conducted an interview study of participants’ perceptions of nonverbal communication in social VR. The authors uncovered several types of nonverbal communication methods used in social VR, including applauding, dancing, and even flying or using emojis. Among the most common across platforms were hand gestures and head movements, like waving or nodding. Similarly, Tanenbaum et al. \cite{tanenbaum2020make} explored ten popular social VR platforms and took inventory of their existing designs for expressing nonverbal behaviors. While the quality of each platform’s support of these behaviors varied, they found that most social VR platforms included the ability to express behaviors such as proxemics, facial expressions, and gestures.

Other researchers have explored particular types of nonverbal behaviors, including gestures, nodding, and eye gaze, and how they affect perceptions of conversation partners in VR \cite{ide2020effects, aburumman2022nonverbal, kurzweg2021using, kevin2018virtual, llobera2010proxemics, yee2007unbearable}. For instance, Aburumman et al. \cite{aburumman2022nonverbal} conducted VR studies with 21 participants, having participants experience two types of nodding behavior in a conversation with agent-avatars. They found that an agent-avatar that nodded while participants spoke resulted in higher levels of trust felt towards the agent-avatar. Ide et al. \cite{ide2020effects} observed the effect of symbolic gestures to invoke gestures on virtual avatars (e.g., thumbs-up emoji to do a thumbs-up, surprised emoji to make a surprised face) on brainstorming tasks. They found that symbolic gestures helped participants express their emotions and supported social communication. Similarly, Kurzweg et al. \cite{kurzweg2021using} explored the effect of body language (e.g., crossing arms, drinking) on communication in VR and found that body language can indicate willingness to communicate and attentiveness towards others in a virtual environment.

These investigations demonstrate the importance of nonverbal behaviors for communication in VR. Despite this, there has been little exploration of making these behaviors accessible to support BLV communication needs \cite{segal2024social, wieland2022non, collins2023making, ji2022vrbubble, wieland2023vr}. One of the most notable efforts is Wieland et al.’s \cite{wieland2022non} work on identifying important nonverbal behaviors in social VR. Wieland et al. conducted interview studies with eleven participants, seven BLV and four sighted companions of BLV people, to identify which nonverbal behaviors they used most often in conversation, and which should be carried over into VR for effective communication. They found that gaze, head direction, head movements, and facial expressions were all important to identify, though they could be difficult to notice for BLV participants. 

Other researchers have developed design guidelines for making certain nonverbal behaviors accessible. Collins and Jung et al. \cite{collins2023making} probed the design space of accessible gaze feedback in VR with a blind co-designer. The co-designer experienced and adjusted feedback for two kinds of gaze (mutual and resting gaze) through 5 parameters (e.g., Modality, Strength, Duration, etc.) to create their preferred gaze feedback in a VR prototype. Their study concluded with recommendations to send haptic vibrations to a blind user via handheld controllers when someone was looking at them in VR. Ji et al. \cite{ji2022vrbubble} explored avatar proxemics in VR. They designed a system where audio beacons indicated proxemic information and tested it in user studies with 12 BLV participants. The system improved BLV people’s awareness of others, providing concrete design recommendations for representing proxemics in social VR.

Both of these works have introduced initial approaches to making nonverbal behaviors accessible to BLV people and are an important step in supporting communication. However, they each only focus on one type of nonverbal behavior at a time, namely eye gaze and avatar proxemics. People use multiple nonverbal behaviors (e.g., eye gaze, head movements, and gestures) when engaging in a social space. As stated by Wieland et al. \cite{wieland2022non, wieland2023vr}, BLV people want access to multiple nonverbal behaviors that are used in conversation. In order to effectively support BLV people’s conversation needs, multiple nonverbal behaviors should be made accessible at once. In addition, these works conducted their user evaluations with pre-recorded agent-avatars in scripted conversations, rather than supporting real conversations with unpredictable human conversation partners. These limitations represent a significant gap in current research efforts to support communication in VR for BLV people. Our work seeks to address this gap by implementing multiple nonverbal cues in conversations in VR to evaluate how these accessible representations of nonverbal behaviors support real conversations with others.
\section{Designing Accessible Nonverbal Cues}
Our goal was to design a set of robust accessible nonverbal cues that could support conversations in VR. To achieve this, we conducted an iterative user-centered design process to convey a set of nonverbal behaviors via accessible nonverbal cues. We worked in a mixed-ability team with professionals experienced in designing accessible technology. We also conducted a series of early informal design sessions with BLV co-designers. Throughout the design process, we met weekly to discuss and test prototypes.

To guide our design process, we developed four design considerations based on: (1) discussions among our mixed-ability team, (2) informal design sessions with BLV co-designers, and (3) past work in designing accessible representations of nonverbal cues in VR \cite{collins2023making, ji2022vrbubble}.

\begin{itemize}
\item\textbf{Cues should leverage nonvisual modalities supported by VR technology.} We should consider the nonvisual capabilities of commercial VR technologies such as haptic vibrations, audio patterns, and spatialized audio.

\item\textbf{Audio cues should represent the emotions behind each nonverbal behavior.} Leveraging the versatility of audio, audio cues should evoke emotions that match the behaviors they are representing, like positive emotions for a smile.

\item\textbf{Cues should be unobtrusive during a conversation.} Cues should not be too loud, long, or distracting to the point where they interrupt or make conversation difficult for users.

\item\textbf{Cues should be understandable when multiple cues are being used simultaneously.} Since people use multiple nonverbal behaviors simultaneously in conversation, the cues should be playable simultaneously, without overwhelming the user.
\end{itemize}

After finalizing these design considerations, we needed to establish which nonverbal behaviors to make accessible. We first considered the broad scope of possible nonverbal behaviors identified by prior work and narrowed them down to a set that is useful in conversations for BLV people. Wieland et al. \cite{wieland2022non, wieland2023vr} found that eye gaze, head direction, head movements, and facial expressions were important for BLV people during conversations. When considering different forms of eye gaze, Collins and Jung et al. \cite{collins2023making}  and Wieland et al. \cite{wieland2022non, wieland2023vr}  found that eye contact information was the most preferred and important for BLV people. Maloney et al. \cite{maloney2020talking} noted that some of the most commonly used nonverbal behaviors in VR included nodding and head shaking. Finally, existing research on facial expressions lists smiling and frowning as two of the most common expressions people make \cite{six-universal-expressions}. Considering these works, we narrowed our focus to five nonverbal behaviors to design as accessible cues: eye contact, head nodding, head shaking, smiling, and frowning.
\subsection{Design Process}
To create our initial designs, we examined existing sound libraries used to indicate information. After sorting through a variety of libraries (e.g., Facebook's emoji sound effects \cite{facebook-soundmoji}, vocal bursts such as laughs and sighs \cite{parsons2014introducing}) we created our initial designs of cues for the five nonverbal behaviors using Paquette et al.’s database of musical emotional bursts. Each burst in this dataset was a brief music clip that Paquette et al.’s participants associated with specific emotions \cite{paquette2013musical}.

We sought to iterate on the initial designs of our cues with BLV users to ensure they would be usable and understandable. To do this, we conducted studies with six BLV participants. Each participant took part in a single in-person 90-minute session where they experienced the most recent iteration of our cue designs. We wanted to demonstrate our cue designs to participants in various scenarios, both in the physical world and in VR, to give them a good idea of what having conversations with these cues would be like. Thus, each study session contained four parts: (1) an introduction to the nonverbal cues in the physical world, (2) a one-on-one conversation augmented by the cues in the physical world, (3) a one-on-one conversation augmented by the cues in VR, and (4) a three-person group conversation in VR.

We represented the cues for participants in different ways in the non-VR and VR settings. In our non-VR introduction to the cues, we played each of the cues (eye gaze, head nodding, head shaking, smiling, or frowning) one by one from a laptop to share the cue audio while participants held VR controllers to receive haptic feedback. In the non-VR conversation, participants again held controllers to receive haptic feedback, but the researcher who was speaking to them played the audio cues from a phone in their hand, so that participants would hear the audio coming from the direction of the person they were speaking with. Finally, for the VR settings, we developed a VR prototype where researchers and the participant would enter a multi-user virtual environment together. Within this environment, the researchers could manually trigger nonverbal cues when they performed certain nonverbal behaviors to augment conversations. Participants used a VR headset and controllers to enter the scene and heard audio cues spatialized to the researchers’ avatars while feeling haptic feedback from cues in their controllers.

We wanted to improve our cue designs with participants and allow them to try any suggestions they had for the cues during their study session. To do this, we asked participants for their perceptions of each cue after the conversations, including emotion invoked, level of distraction, and suggestions for improvements. After discussing the initial cue designs, we worked with participants to develop new iterations of the cues based on their critiques. Participants could request changes such as the length of the haptic cue, the volume of the audio, and the type of audio file being played (e.g., a musical note, a TV show sound effect, a recorded laugh, etc.). Participants could also create their own audio cues from scratch by directing the research team to create new sound effects. After making changes, we would demonstrate the new cues for participants, prompt them for feedback, and continue iterating. All of the changes we made to the cues were implemented via a laptop running Unity and the audio mixing software Garageband. Participants experienced the new cue iterations by listening to audio played from the laptop and holding VR controllers to receive haptic feedback. At the end of the session, we had participants comment on which iterations of the cues they preferred the most. We reached consensus for final designs when three participants in a row preferred the current versions of the cue designs without requesting modifications.

Following these sessions, we developed three additional design considerations from participants’ feedback:

\begin{itemize}
\item\textbf{Accessible audio cues should be short for usability and accuracy.} Shorter lengths for the accessible nonverbal cues were preferred. If cues are too long, they become disruptive to the conversation (e.g., a two-second clip representing head shaking drowned out conversation and tended to last over a second after head movements stopped).
\item\textbf{Familiar sound effects are preferable for inciting emotions.} Sound effects that imitated sounds from media sources, like television game shows, were easier to associate with specific emotions than musical patterns.
\item\textbf{Haptics should be used to represent frequent nonverbal cues.} Since audio repeating continuously would disrupt conversation, cues such as eye contact that occur frequently should be represented by less-disruptive haptic vibrations. 
\end{itemize}

\subsection{Final Designs}
After receiving consistent feedback from participants that cues were easy to understand and unobtrusive to conversation, we stopped iterating on our designs. The final designs were as follows \footnote{All tracks for audio cues are provided here: \url{https://soundcloud.com/shadowdios/sets/nvc-study-sound-effects}}:

\begin{itemize}
\item\textbf{Eye Contact.} A continuous haptic buzz that lasted for the duration of the eye contact.
\item\textbf{Head Nod.} A succession of one high-pitched note followed by a slightly lower note from a xylophone, repeating twice and lasting 1 second.  
\item\textbf{Head Shake.} A succession of one low-pitched note followed by a slightly lower note from a flute, repeating twice and lasting under 1 second. 
\item\textbf{Smile.} A high-tone chime-like sound effect, formed of a series of cheerful ringing notes lasting 2 seconds.
\item\textbf{Frown.} A low-tone trumpet sound effect lasting for 1 second. 
\end{itemize}

An overarching takeaway we found with these cues was that \textbf{participants found a set of five cues was overwhelming to learn and immediately use in conversation}. Thus, we decided to \textbf{explore a smaller set} of nonverbal cues, selected from these five. We selected eye gaze for this subset since our participants responded the most positively to eye contact and its non-intrusive, haptic form. We also chose the two head movement cues–head nodding and head shaking–since they are more commonly used in mainstream VR than facial expressions due lack of integration with face-tracking technology.
\subsection{Discussion and Implications}
Our goal for the design process was to iteratively design accessible versions of nonverbal cues. We had not yet tested how effective these designs were at conveying social information about a conversation partner. One type of social information that many of our participants were particularly interested in was attention, specifically, in “reading the room” to know if their partner was paying attention to them. However, as we had focused primarily on determining how disruptive or intuitive the cue designs were, we had not examined the cues’ abilities to convey information like this. Further work was required to specifically evaluate how well cues can convey a conversation partner’s level of attention.

We had also only tested the cues with a small set of co-designers who were used to experimenting with and designing new accessible technologies. While their feedback was useful for designing effective nonverbal cues, it did not represent how easy these cues would be to learn and utilize for other BLV users. To truly understand how useful our cues would be, we needed to evaluate them with a larger and more diverse group of BLV participants.

\section{Evaluating the Cues in Conversations}
We aimed to evaluate the cues’ effectiveness in real-time conversations in VR. To do this, we needed to consider what it meant for nonverbal behaviors to support conversations. Nonverbal behaviors convey a wide range of social information, from emotional reactions to level of agreement. For example, someone may look at you to indicate that they are listening or nod their head in agreement with what you are saying. Our previous design process had established our participants’ interest in determining the attention levels of their partners. Since our set of three nonverbal cues, eye contact, head nodding, and head shaking, also commonly represent attention \cite{lawson2015just, eye-contact-communication, attentive-body-language, body-language-listeners}, we selected attention as our focus for the evaluation. This would allow us to see if our cues could support our prior participants’ desired conversational needs and whether our cues could convey important social information to support conversations.

We designed our evaluation with certain key questions in mind:

\begin{itemize}
\item How \textit{accurate} are participants at detecting attention with and without accessible nonverbal cues?
\item How \textit{confident} do participants feel about detecting attention with and without accessible nonverbal cues?
\end{itemize}

\begin{table*}
\centering
\caption{\label{tab: Participant Demographics} Participant demographics. F=Female; M=Male; All vision measures (acuity, visual field, light perception) were self-reported by participants.}
\begin{tblr}{
  width = \linewidth,
  colspec={Q[70]Q[70]Q[200]Q[300]Q[400]Q[500]Q[500]},
  hlines,
  vlines,
}
\textbf{P\#}        
& \textbf{Age}
& \textbf{Gender}     
& \textbf{Vision}                                                                  
& \textbf{Visual Acuity}
& \textbf{Visual Field}
& \textbf{Light Perception}

\\

P1       
& 37                                                          
& F
& {Low vision\\Blind}
& {R: 20/320\\L: 20/160}
& {R: no peripheral\\L: wider visual field}
& Yes

\\

P2   
& 51
& F
& Low vision
& {R: 20/700\\L: 20/1200}
& {R: limited peripheral\\L: ok}
& Yes 

\\

P3       
& 67
& M
& Low vision
& {R: Unsure\\L: 20/200}
& {R: Unsure\\L: full}
& Yes, left eye                                               

\\

P4        
& 27
& M
& Blind
& Unsure
& Unsure
& Yes

\\

P5
& 21
& M
& Low vision
& Unsure
& None
& Yes, right eye  

\\

P6
& 18
& F
& Blind
& {R: blurry,\\2-3 inches\\L: No peripheral,\\23/200}
& Unsure
& Yes

\\

P7
& 50
& M
& Blind
& None
& None
& None

\\

P8
& 36
& M
& {Low vision\\Blind}
& Unsure
& Lacks central, left eye is slightly better 
& Limited on the peripheral

\\

P9
& 45
& M
& Blind
& None
& None
& None

\\

P10
& 74
& F
& Blind
& None
& None
& None

\\

P11
& 26
& M
& {Low vision\\Blind}
& {R: 20/300\\L: 20/400}
& Slight central in left eye, 180 degree visual fields
& Yes

\\

P12
& 66
& M
& {R: Low vision\\L: Blind}
& Unsure
& Unsure
& Yes

\\

P13
& 34
& M
& Low vision
& Unsure
& Unsure
& Yes

\\

P14
& 45
& F
& Blind
& None
& None
& None

\\

P15
& 55
& M
& Low vision
& {R: Unsure\\L: Unsure}
& Unsure
& Yes, left eye

\\

P16
& 43
& M
& Blind
& None
& None
& None

\end{tblr}
\end{table*}

\subsection{Methods}
We recruited 16 participants from an organization that provides services to BLV people. The participants identified as blind or low vision (5 females, 11 males) whose ages ranged from 18 to 74 (mean = 43.44, standard deviation (SD) = 44, Table 1).

Each participant took part in a single in-person 90-minute session. One researcher led the study and two researchers participated in the study as conversationalists in the virtual environment. During the study session, participants first completed a tutorial on the VR system and accessible cues. Then, they completed three conversation tasks in virtual environments: (1) a practice task, (2) a baseline task, and (3) a treatment task. The researchers would introduce a new conversation topic at the beginning of each conversation task, including the practice task. We counterbalanced the order of the baseline and treatment tasks. We concluded the session with a semi-structured interview. 

\subsubsection{\textbf{VR Prototype.}} We developed a VR prototype with multi-user virtual environments to give participants a virtual conversation space to test the cues in. These environments allowed participants and researchers acting as conversationalists to join the space and talk together. We designed the prototype to automatically detect when the researchers nodded their heads, shook their heads, or made eye contact with the participant. This automatic detection allowed the researchers to focus on the conversation tasks and their assigned attention conditions (see section \ref{Counterbalanced Baseline and Treatment Tasks}) instead of manually triggering cues. Since participants did not need access to their own nonverbal cues, participants’ cues were ignored by the system to prevent confusion. The prototype detected head movements by noting repeated translations of the VR headset left and right (shaking head) or up and down (nodding). It detected eye contact using virtual raycasts sent from a “hit box” around the eye levels of each avatar; if these raycasts both reached each other’s hitboxes, the prototype determined eye contact was being made. If any of these behaviors were detected, the prototype triggered the accessible cues, playing the corresponding audio or haptic feedback on the participants’ headset and controllers.

The prototype played all nonverbal cues at once for any behaviors that were detected from the researchers who were in the virtual space. This meant participants often received feedback for accessible cues from the researchers at the same time. To determine who the cues were coming from, participants used a combination of audio spatialization and their own head movements in VR. All audio cues were spatialized so participants heard cues coming from the researcher in the room triggering them. For eye contact, our haptic cue, participants received haptic feedback when they were oriented toward the researcher who was looking at them, i.e., if they felt haptic vibrations when looking left, they knew the researcher on their left was the one making eye contact with them. In this way, participants were easily able to distinguish between different researchers’ cues during conversation.

\begin{figure}
\includegraphics[width=250pt]{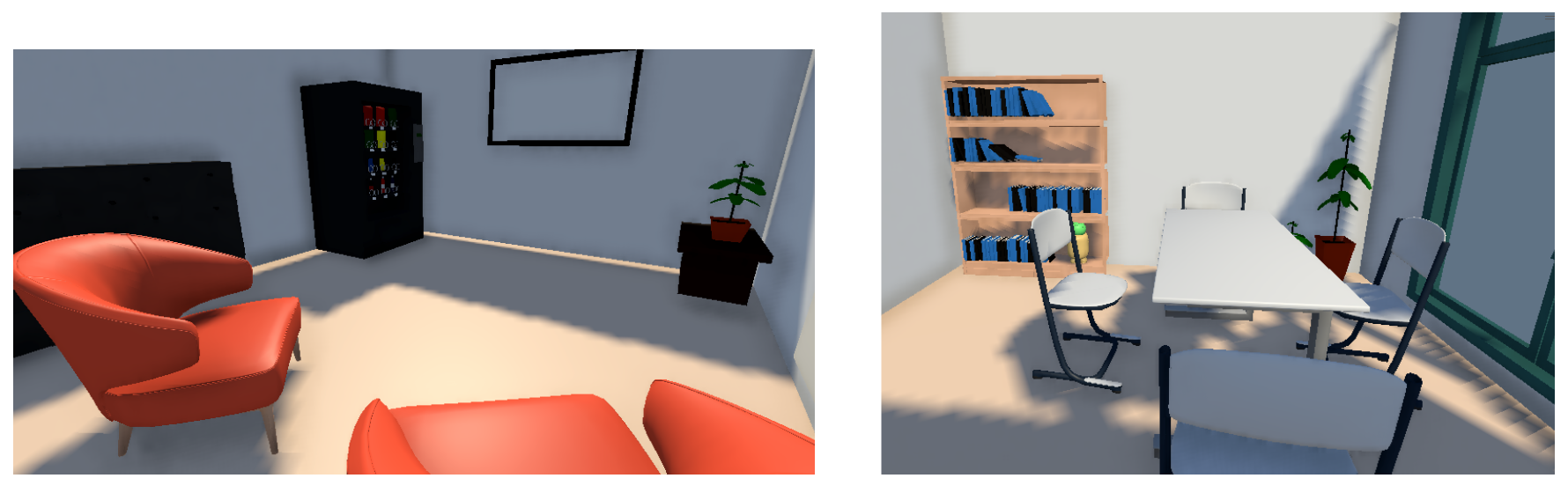}
\caption{\label{fig: Scenes in Protoytpe.} The virtual environments used in the study. Left: the tutorial room, designed to look like a small waiting room. Right: the task room, designed to look like a professional conference room. The same room was used for both baseline and treatment tasks.}
\end{figure}
%% Accessible Description: Two images of the virtual environments that the studies took place in. The left image is a screenshot of a small, white room designed to look like a waiting room. Two orange chairs are placed facing slightly towards each other, and the screenshot is taken as though the viewer is seated in one of the chairs. There is a cabinet against the left wall, a vending machine in the left corner of the room, and a potted plant in the right corner. A TV hangs on the back wall. The right image is a screenshot of a slightly larger white room designed to look like a conference room. There is a long white table with four white chairs placed around it. The screenshot is taken as though the viewer is standing behind one chair. There is a bookshelf against the far wall, a potted plant in the right corner across from the bookshelf, and a large window taking up the right wall.

\subsubsection{\textbf{VR Tutorial.}} We began the session with a ten-minute VR equipment tutorial and introduction to our accessible nonverbal cues. Then, we explained how to wear a VR headset and use the controllers. We also explained and played each of the three accessible nonverbal cues.

\subsubsection{\textbf{Practice Task.}} Participants entered the virtual environment and engaged in a ten-minute conversation with two researchers using the cues. The topic was how things have improved after the pandemic. We split the task into two segments, pausing the conversation after five minutes. This allowed us to conduct checks on participants’ thoughts towards the start of the conversation, and towards the end when they were more familiar with their conversation partners. After each segment, we asked three questions:

\begin{itemize}
\item Can you describe what you just experienced in the virtual space?
\item What were the reactions of the other speakers?
\item “What’s your level of confidence in your answer to the previous question on a scale of 1 to 5, where 1 is not at all confident and 5 is very confident?”
\end{itemize}

Finally, we asked a set of Likert-scale questions and an open-ended question to allow participants to reflect on the experience. 

\subsubsection{\textbf{Counterbalanced Baseline and Treatment Tasks.}}\label{Counterbalanced Baseline and Treatment Tasks} Participants experienced a 15-minute debate task. With this task, we wanted to create a situation where the participant would need to figure out whether attention was on them or not without the researchers explicitly saying anything about their attention state. Thus, we designed the task as a debate since it would require that researchers first pose topics and then fall silent to listen to the participant’s response. Two researchers acted as conversation partners for the participant during the debates. The researchers would present a casual topic for debate, such as whether hot dogs were considered a sandwich, and then prompt the participant for their position. Following this, the participants’ goal was to convince their conversation partners of their position on the topic. The participants completed this task twice in a counterbalanced order: once with the cues (treatment task) and once without (baseline). They were presented with a new topic for each debate.

\begin{table*}
\centering
\caption{\label{tab: Tasks Table} A table showing the four different task sets, with two counterbalanced attention conditions and one debate topic per each task. }
\begin{tblr}{
  width = \linewidth,
  colspec={Q[30]Q[200]Q[200]Q[200]},
  hlines,
  vlines,
}
\textbf{Set}        
& \textbf{Debate Topic}
& \textbf{Segment 1 Attention Condition}     
& \textbf{Segment 2 Attention Condition}
\\

A       
& if a hotdog is a sandwich
& Nobody pays attention
& Only Researcher A pays attention

\\

B   
& morning or evening showers are better
& Everyone pays attention
& Only Researcher B pays attention

\\

C       
& if pineapple belongs on pizza
& Only Researcher B pays attention
& Only Researcher A pays attention                                    

\\

D        
& if sandwiches should have crust or no crust
& Everyone pays attention
& Nobody pays attention

\end{tblr}
\end{table*}

\textbf{Attention Conditions.}  We created four “attention conditions” as a ground truth of the researchers’ attention states (whether the researchers were paying attention to the participant at a given time). Participants were not given these attention conditions; these were for the researchers only. To determine what it meant for a researcher to be “paying attention,” we looked at standard definitions of attention from psychology. One of the most commonly used definitions of attention in prior work comes from James’ \textit{Principles of Psychology}, which defines attention as the “[f]ocalization, concentration, of consciousness” on “one out of what seem several simultaneously possible objects or trains of thought” \cite{james2007principles}. In other words, “attention” is when a person focuses on one object out of several others. To operationalize this, we had our researchers focus on the participant and \textit{demonstrate} their focus by using verbal and nonverbal cues to respond to whatever the participant was saying in conversation. If they responded to the participants’ thoughts and behaviors for over half the duration of the conversation, we considered the researcher to have been focused on the participant for the majority of the conversation. Thus, they were “paying attention” to the participant by standard definitions from psychology.

In total, we had four attention conditions: (1) \textbf{both researchers} were paying attention to the participant, (2) \textbf{neither researcher} was paying attention to the participant, (3) \textbf{only researcher A} was paying attention to the participant, and (4) \textbf{only researcher B} was paying attention to the participant.

The 15-minute debate task was split into two 7.5-minute segments in which participants experienced \textbf{one} attention condition per segment. Since participants repeated this task twice, once for the baseline and once for the treatment, they experienced \textbf{four} attention conditions in total. The order in which they experienced the attention conditions was counterbalanced (see table \ref{tab: Tasks Table}). After each segment, we asked the following: 

\begin{itemize}
\item(1) Can you describe what you just experienced in the virtual space? 
\item(2) Did \textbf{researcher A} pay attention to you? 
\item(3) Did \textbf{researcher B} pay attention to you? 
\item(4) “What’s your level of confidence in your answer to the previous question on a scale of 1 to 5, where 1 is not at all confident and 5 is very confident?” 
\end{itemize}

This was meant to gauge participants’ confidence level in their ability to determine the attention states of their conversation partners.

At the end of each task, we asked Likert-scale questions to help us understand participants’ experiences with and without cues. We asked participants to rate their agreement with the following statements about the quality of their experience:

\begin{itemize}
\item “The accessible nonverbal cues were useful to detect if people were paying attention to me or not”
\item “The accessible nonverbal cues were distracting”
\item “It was easy to determine whether people were paying attention to me or not”
\end{itemize}

Additionally, we asked participants to rate their agreement with the following statement: “It was easy for me to determine whether attention was on me or not.” After the treatment task, we asked them to rate their agreement with the following: “The accessible nonverbal cues were distracting” and “The accessible nonverbal cues were useful to detect attention.” 

\subsubsection{\textbf{Semi-Structured Interview.}} We ended the study with a 20-minute semi-structured interview, in which we asked the participants to reflect on their experience and preferences for accessible nonverbal cues. We discussed scenarios where cues may be useful, other nonverbal behaviors, other effects of the cues, and suggestions to improve the cues.

\subsubsection{\textbf{Data and Qualitative Analysis}} All participants completed all of the tasks and interviews. Audio and video recordings from the study session were collected and transcribed using an automatic transcription service, Otter.ai. Two researchers coded the transcripts using open descriptive codes. They coded the same two transcripts, then came together and discussed discrepancies. Through discussion, they generated a codebook and split the rest of the data. Afterward, they conducted a thematic analysis \cite{braun2006using} using affinity diagrams to categorize the codes into themes. 

\subsubsection{\textbf{Statistical Analysis.}} We also conducted a statistical analysis to identify whether the nonverbal cues had a significant effect on participants' accuracy and confidence when assessing attention states. We used a linear mixed effect model fitted by Restricted Maximum Likely estimation to model our data. We used this type of model since we had collected repeated measures of participants’ reactions to the conversation at varying segments, as well as to handle cases where participants had refused to answer our measures, creating occasional missing data points. \textit{ParticipantID} was used as a random effect to indicate to our model that multiple measures were attached to each Participant ID (i.e., that we collected multiple measures from our participants). T-tests run on the models used Satterwaithe’s method to account for variances in our data samples.

\textbf{Effect of Nonverbal Cues on Accuracy and Confidence.} We explored the effect of the nonverbal cues on participants’ accuracy and confidence in assessing attention states. To do this, we fit one model with one factor, \textit{Group} (two levels: \textit{withCues, withoutCues}), one measure for accuracy, \textit{Accuracy}, and one measure for confidence, \textit{Confidence}. The values for \textit{Group} were pulled from the study design, whether users were experiencing the scenario with or without the cues. The \textit{Accuracy} and \textit{Confidence} scores were pulled from participants' assessments of attention states during each scenario segment. Each participant provided four assessments per task \textit{of their opinion} of whether a particular speaker was paying attention: one for speaker A in segment 1, one for speaker A in segment 2, one for speaker B in segment 1, and one for speaker B in segment 2. We had 16 participants who each provided eight assessments, resulting in 128 total assessments.
\subsection{Findings}
\subsubsection{\textbf{Task Performance and Statistical Tests.}}
We report participants’ accuracy and confidence when assessing the attention states of their speaking partners. We also report statistical tests to identify whether having the nonverbal cues had a significant effect on participants’ accuracy and confidence. We quantified the usefulness of the cues by examining the impact of one factor, \textit{Group}, with two levels (\textit{withCues} and \textit{withoutCues}) on participants’ performance. We interpreted “significance” as a p-value < 0.05, established prior to running tests.

\begin{figure}
\includegraphics[width=250pt]{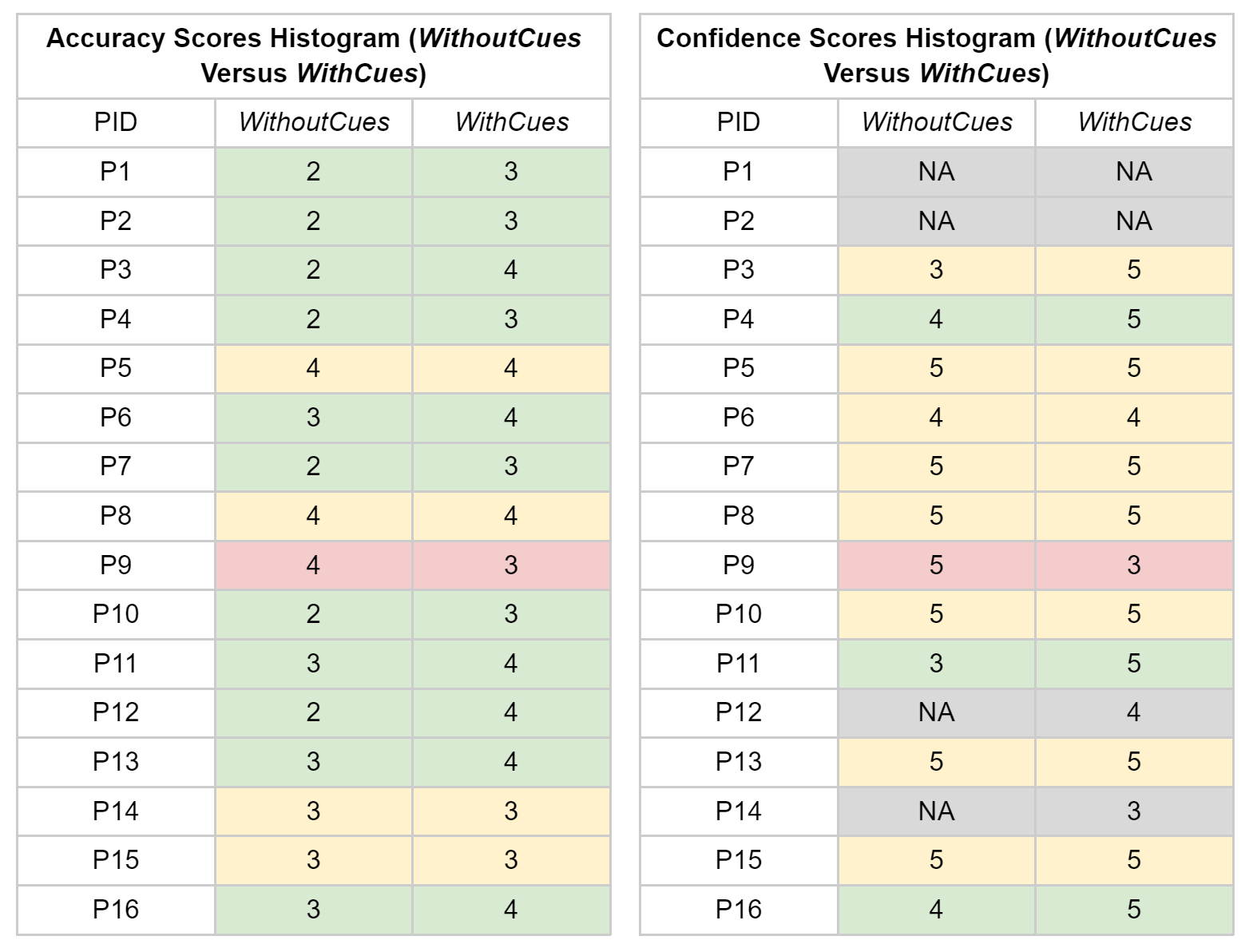}
\caption{\label{fig: Task Performance Histograms.} Histograms comparing each participants’ accuracy and confidence scores between the task without cues and the task with cues. Red indicates worse scores with cues, yellow indicates equal scores, and green indicates better scores with cues. Gray indicates one or more of the answers were not provided by participants (NA) so a comparison cannot be made.}
\end{figure}

\textbf{Accuracy.} We report scores for 16 participants (see figure \ref{fig: Task Performance Histograms.}). 11 out of 16 participants had better accuracy scores when using the accessible nonverbal cues. Out of the remaining five, four participants scored the same as they did without cues, and one participant scored marginally better (4 correct without cues versus 3 correct with cues). 

%% Accessible Description: Two tables depicting the accuracy and confidence scores in a histogram. The left table is titled "Accuracy Scores Histogram (WithoutCues Versus WithCues)". It has three columns: PID, WithoutCues, and WithCues. Under PID, all participant ID numbers are listed from P1 to P16, with each number being assigned a row. Under WithoutCues and WithCues, the participants' accuracy score for each condition is displayed, with colors to indicate if the scores improved WithCues. Almost every row is green, with a few yellow rows scattered in, and one red row. The right table is titled "Confidence Scores Histogram (WithoutCues Versus WithCues)". It is set up with the exact same columns as the previous table. This table has mostly yellow rows, with a few green rows scattered in, and one more red row. In addition, there are four gray rows where participants' scores are written as NA.

\textbf{Effect of Nonverbal Cues on Accuracy.} We found a significant effect of \textit{withCues} on \textit{Accuracy} (t=2.614, p=0.010). The mean accuracy of participants across all conversation scenarios while using the cues was 87.5\% (SE=0.1), compared to 68.8\% accuracy (SE=0.1) without the cues. We conclude that participants’ accuracy was significantly higher when using the cues. Thus, the cues helped their ability to accurately judge how much attention people were paying to them during conversation.

\textbf{Confidence.} We report scores for 12 participants (see figure \ref{fig: Task Performance Histograms.}). The remaining four did not provide confidence scores. Half the participants, six out of 12, were more confident when using the accessible nonverbal cues. Out of the remaining six, three of them scored their confidence the same as they did without cues, and the other three reported higher confidence without cues. However, these three participants all tended to score their confidence very high across all conditions, having an average score of 17.7 for confidence out of a maximum confidence score of 20 across all conditions.

\textbf{Effect of Nonverbal Cues on Confidence.} We found a significant effect of \textit{withCues} on \textit{Confidence} (t=2.224, p=0.028). The mean confidence participants reported across all conversation scenarios while using the cues was 4.0 (SE=0.4) out of 5, compared to a confidence score of 3.7 (SE=0.4) out of 5 without the cues. We conclude that participants’ confidence was significantly higher when using the cues. Thus, the cues improved how confident participants felt that they knew how much attention people were paying to them, and so they were more willing to act on their judgments of attention.

\textbf{Quality of Experience.} Participants provided scores on a scale of 1 to 5, where 1 meant they strongly disagreed and 5 they strongly agreed with a given statement about the quality of their experience with the cues. Each statement, also reported in section \ref{Counterbalanced Baseline and Treatment Tasks}, related to a quality metric as follows:

\begin{itemize}
\item{[Usefulness] “The accessible nonverbal cues were useful to detect if people were paying attention to me or not”}
\item{[Distraction]  “The accessible nonverbal cues were distracting”}
\item{[Task Ease] “It was easy to determine whether people were paying attention to me or not”}
\end{itemize}

\begin{figure}
\includegraphics[width=250pt]{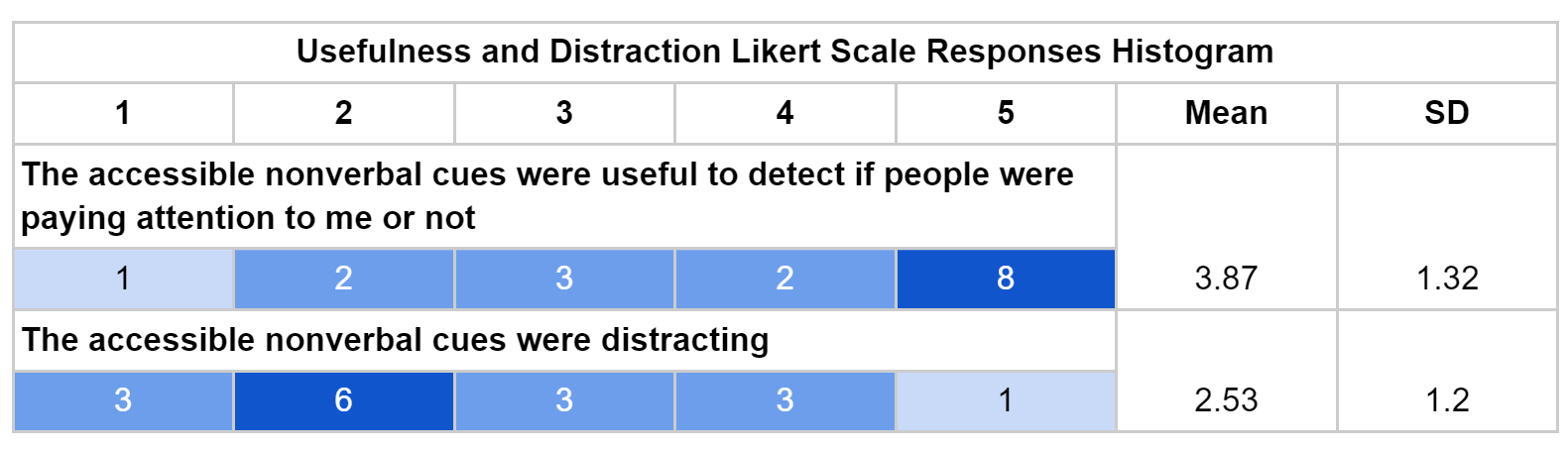}
\caption{\label{fig: Quality 1 Usefulness Distraction Histograms.} Likert-scale score responses for cue usefulness and distraction. }
\end{figure}

\textbf{Usefulness.} Scoring usefulness helped us understand whether participants found the cues useful, regardless of other difficulties with the cues. Participants reported high scores for usefulness (see figure \ref{fig: Quality 1 Usefulness Distraction Histograms.}). 10 out of 16 participants agreed or strongly agreed the nonverbal cues were useful, and three neither agreed nor disagreed. Interestingly, the three participants who disagreed or strongly disagreed had low vision. These results indicate that most participants found the cues useful, but low vision people might find them less useful than blind people (possibly due to their residual vision).

\textbf{Distraction.} Scoring distraction helped us understand whether participants felt distracted from their conversation with the other speakers when using cues. Participants reported relatively low scores for distraction (see figure \ref{fig: Quality 1 Usefulness Distraction Histograms.}). Nine out of 16 participants disagreed or strongly disagreed, and three participants neither agreed nor disagreed. The four participants who agreed or strongly agreed had low vision, as seen above with usefulness. These results indicate that most participants did not find the nonverbal cues distracting, but that low vision people might find them more distracting than blind people (possibly due to having to split their attention between their residual vision, the cues, and the speakers’ speech). 

%% Accessible Description: A histogram depicting Likert scale responses from participants. The histogram is titled "Usefulness and Distraction Likert Scale Responses Histogram". There are five columns labled 1 through 5, plus two more columns labled "Mean" and "SD". Under the numbers, there are statements listed that participants responded to. Underneath each statement, there are numbers indicating how many participants responded to the statement with the given column number (e.g., if 8 people responded with a score of 5, the "5" column would have the number 8 under the statement). Further, the numbers are color-coded in shades of blue to indicate where the most amount of answers are falling: the deeper blue colors mean more responses, and lighter blues indicate fewer responses. The first statement is "The accessible nonverbal cues were useful to detect if people were paying attention to me or not." The answers are concentrated towards 5, with this column being a very dark blue, and 1-4 being light shades of blue. "Mean" is listed as 3.87 and "SD" is 1.32. The second statement is "The accessible nonverbal cues were distracting." The answers are concentrated towards 2, with this column being a very dark blue, and columns 1, 3, 4, 5 being light shades of blue. "Mean" is listed as 2.53 and "SD" is 1.2.

\begin{figure}
\includegraphics[width=250pt]{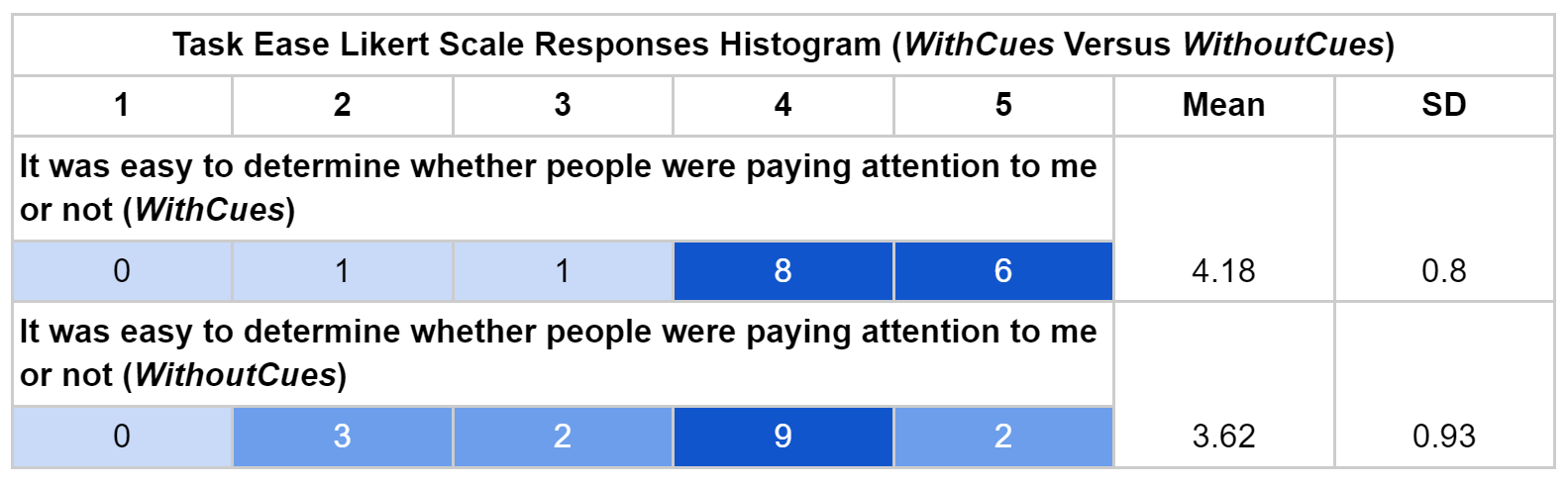}
\caption{\label{fig: Quality 2 Task Ease Histograms.} Likert-scale score responses for task ease with cues and without cues.}
\end{figure}

\textbf{Task Ease.} We compared the metric of task ease between the \textit{withCues} and \textit{withoutCues} conditions. This helped us understand how much effort participants felt they had to put in to determine attention states. Participants generally reported higher scores for task ease with cues (see figure \ref{fig: Quality 2 Task Ease Histograms.}). Nine out of 16 participants reported higher ease scores with cues than without cues, three reported the same scores with and without cues, and four reported lower scores with the cues. These results indicate that most participants felt it took less effort to determine attention states when the cues were present, though a small minority found it more challenging.

%% Accessible Description: A histogram depicting Likert scale responses from participants. The histogram is titled "Task Ease Likert Scale Responses Histogram (WithCues Versus WithoutCues)". There are five columns labled 1 through 5, plus two more columns labled "Mean" and "SD". Under the numbers, there are statements listed that participants responded to, with parantheses indicating whether this statement was for the WithCues or WithoutCues condition. Underneath each statement, there are numbers indicating how many participants responded to the statement with the given column number (e.g., if 6 people responded with a score of 5, the "5" column would have the number 6 under the statement). Further, the numbers are color-coded in shades of blue to indicate where the most amount of answers are falling: the deeper blue colors mean more responses, and lighter blues indicate fewer responses. The first statement is "It was easy to determine whether people were paying attention to me or not (WithCues)." The answers are concentrated towards 4 and 5, with these column being very dark blue, and 1-3 being light shades of blue. "Mean" is listed as 4.18 and "SD" is 0.8. The second statement is "It was easy to determine whether people were paying attention to me or not (WithoutCues)." The answers are concentrated towards 4, with this column being a very dark blue, and columns 1, 2, 3, 5 being light shades of blue. "Mean" is listed as 3.62 and "SD" is 0.93.

\subsubsection{\textbf{Participants’ Reactions to the Cues}}
We report participants’ reactions to the cues in terms of their usefulness, usability, and emotional impact, three metrics often used when evaluating user experience \cite{usability-usefulness-ux, emotional-impact-ux}.

\textbf{Usefulness.} Most participants felt the cues were useful, allowing them to determine others’ behaviors in ways they normally could not. For example, the cues helped P9 determine if people were ignoring him in the conversation:  “I feel like there was less attention being paid [to me]...I was hearing fewer of the head nods and shakes and making eye contact was a little more difficult, too.” P11 used the cues to reassure himself that others had been paying attention to him: “They were…engaging with my points that I brought up and nodding or not nodding towards my answers.” Overall, we found that eight participants used the cues to determine attention by connecting the amount of nonverbal cues they noted with the level of attention given to them.

While many participants found the cues useful, others were frustrated by them. Three participants felt they were able to converse adequately without the cues, and found the addition of cues “distracting.” These participants mentioned the cues might be useful in certain scenarios, such as when no one is offering them verbal feedback. However, since they did not experience that in the study, the cues did not feel necessary: “There's always one person to talk to. Maybe if none of them were talking to me then it might be a little different.” (P14) 

Participants also sometimes disagreed with the information the cues provided about the speakers, finding that it conflicted with the information they had determined themselves. We especially noted this among our participants with residual vision, who could see movements from the speakers’ avatars (P15, P4). For example, P4 mentioned he “could see the head turning” when speaking with the other avatars, as well as their basic gestures. As such, he preferred to rely on “body language and [head movements]” that he could see, instead of “believing” (P4) the nonverbal cues.

Finally, a few participants doubted whether the cues were useful outside of the study. One participant mentioned he would not use cues in real life until he knew a person well enough to understand “what their movements mean” and thus decide which cues he should be looking for (P15). Another participant, P1, did not want to receive cues from people when he wasn’t conversing with them:

\begin{quote}
I certainly wouldn't want to wear this in, like, outside settings, like riding on the bus. Apparently, people look at me all the time. Whenever I am on the bus my family tells me, ‘Oh, that person's looking at you,’ and I don't care. I'm just doing what I need to do. [So] this wouldn't be useful for me, like, I'm seeing everybody stare at me that'd make me so self-conscious. I don't know how you guys walk around the world knowing everybody's looking at you. It's kind of freeing to not know that. (P1)
\end{quote}

\textbf{Usability.} Participants gave varying feedback on whether they found the cues easy to learn and use. Twelve participants felt the learning curve was low, and commented that the cues seemed like a natural way to provide information about nonverbal behaviors. P9 said, “You get used to [the cues] pretty quickly because it's just a pretty natural thing, and the sounds really are not overpowering. So they would just become part of the scenery, the…auditory scenery.” Two participants agreed that the cues became part of the conversation’s “background” and felt they did not have to “pay attention that much” (P16) to recall the cues’ meaning and apply it to conversation.

Four participants mentioned that it was difficult to remember what the cues signified. For instance, P15 mentioned he could not recall what the cues meant since he would end up being more worried “about how I look to someone” because of the constant eye gaze feedback. P1 also expressed her frustration with trying to connect what the cues were signaling to the speakers’ behaviors. She explained that “The haptic buzzes [don’t help with] figuring out who's looking at me, both of these [controllers] were just buzzing constantly. So I'm assuming everybody was looking at me” (P1).

\textbf{Emotional Impact.} While most participants did not have strong emotional reactions to the cues, some were moved by how much the cues impacted their confidence during conversation. P2 in particular felt empowered by having the cues: “I felt very confident…because in my real [life], I don't get to understand head nods and eye contact…It was empowering to get that feedback.” She explained that this sense of empowerment was heightened by being able to respond to the feedback she was getting, giving her a sense of control. “I really like the autonomy and the ability to know when you're agreeing or disagreeing, instead of having to wait for someone to tell me…there's something kind of powerful about actually knowing.” P5 agreed with these sentiments, explaining that for him, having the cues “reassures that they're listening to me” and gives him the confidence to keep speaking.
\subsubsection{\textbf{Uses of Cues}}
We now report participants’ reflections on possible uses for the cues.

\textbf{Scenarios.} Our participants gave a wide range of scenarios that they imagined cues could be useful for. These scenarios included:

Intimate or Private Interactions:
\begin{itemize}
\item{Going on first dates (P4)}
\item{Watching shows with friends or family (P14)}
\end{itemize}

Professional Interactions:
\begin{itemize}
\item{Attending virtual conferences (P7)}
\item{Attending or facilitating meetings at work (P7)}
\item{Interviewing for a job (P4)}
\item{Having a one-on-one conversation with your boss (P16)}
\end{itemize}

One of our participants, P6, also brought up a social scenario she encounters frequently in her daily life, where sighted people either assume she can see or forget about her visual impairment. “There are those people that we come across who can forget that we can't see, and they'll just shake their head. Yes or no.” In these scenarios, P6 envisioned the cues could provide clarity by telling her about nonverbal behaviors, and prevent sighted peoples’ misconceptions “that we're ignoring you” by helping her respond to those behaviors on time.

Finally, two of our participants (P16 and P2), who were both educators, felt the cues could be helpful for interactions with students. In P16’s case especially, he felt the cues would help him meet access needs for his students:

\begin{quote}
I'm a special education teacher and some of the students I work with are non-verbal. So they have developmental disabilities. They don't speak. And all they do is gesture or point at something if they want something, and then in that case I don't see what they're looking at or what they're pointing at. So they work with, in the classroom, they have aids…So I ask them, ‘Is he looking at me, or is he looking at something else? Is he looking at the toy?... Is he nodding [or] doing a different gesture?’ So if I can do that with smart glasses or something similar like this virtual headset [with cues]...I think it will help me with my work. (P16)
\end{quote}

\textbf{Being A Better Conversationalist.} Some participants felt the cues made them better conversationalists, and “leveled the playing field” (P2) when speaking with sighted people so they could participate in more ways than usual. For example, P2 felt that the cues allowed her to take a leadership role and guide conversation topics.

\begin{quote}
It was rather empowering to be in [the virtual conversation] and to be able to say, ‘Oh, so you agree, or so you disagree. Tell me more about that.’ I feel like [the cues] almost gave me a stronger position of leadership where I can decide where to go next. (P2)
\end{quote}

Other participants mentioned they felt the cues improved the flow of the conversation. P13 explained that often, people only give him “a head nod or a shake. And I'm just like, ‘Okay, I'll just stare at you until an answer comes, or you ask me…what I'm doing.’” He felt the cues removed that step of waiting awkwardly for clarification. As a result, “the repartee [of the conversation felt] a lot smoother because I can say something that's kind of an open-ended question, and just get a nod or a no, and then continue going instead of having the conversation stop and have the other person have to say, yes or no.”
\subsubsection{\textbf{Design Suggestions}}\label{Design Suggestions}
We now discuss various suggestions participants shared to improve the nonverbal cues. 

\textbf{Modality Suggestions.} Participants had various preferences for the modality (haptic or audio) of the cues. Several participants felt that haptics were less disruptive than audio, commenting that haptics are not “overpowering the conversation or overpowering whoever is talking” (P8) and they “wouldn't distract your engagement to the conversation” (P10). P2 also added, “I like the haptic because it leaves room for my listening to be focused on everything else and what people are saying.” 

However, some participants mentioned that haptics may not be as versatile as audio, where you can use various sounds to differentiate cues. P16 noted that haptics could be difficult to use for multiple nonverbal cues since “we are more limited” to simple vibrations which “might be hard to differentiate” from each other, especially if the system is trying to signal more than one cue at once. To meet this shortcoming, some participants suggested using both haptics and audio to have “multiple ways of disseminating information” (P2) during a conversation.

P9 explained that modality preferences could vary based on user ability and suggested that both haptics and audio should remain to account for this. P9 explained that for “someone who's deaf-blind, having that haptic feedback is a big one that could help them be more engaged in communication. … But, on the other hand, if you have someone who's lost sensitivity in their hands, haptic is not great, either. So I say, keep both paths available” (P9).

\textbf{Suggestions for Additional Cues.} Participants had various suggestions for additional cues. Some requested more gestures including someone looking at their cell phone (P14), raising their hand (P7), or reaching out for a handshake (P13). 

Participants also wanted information about more subtle nonverbal behaviors. P1 was interested in receiving specialized eye gaze information, such as having one cue for concentrated eye contact versus another for a quick glance. P7 said that facial expressions like a “dirty look” or a “look of total disengagement” would be helpful. While most participants were interested in more cues, others brought up concerns that subtle cues may occur too often: “There just seems to be so much more possibilities…that I wouldn't want to get information for, like slightly sad or slightly happy. It seems like an information overload.” (P11)

\section{Discussion}
Addressing our research question, we found our designs of accessible nonverbal cues for eye contact, head nodding, and head shaking effectively supported conversations in VR for BLV users. Moreover, our designs provided participants with a concrete experience of holding real conversations in VR with access to nonverbal behaviors, which allowed them to reflect on possibilities for accessing more forms of nonverbal social information. The design and evaluation of the nonverbal cues allowed us to identify novel design implications for nonverbal cues. 

In the following sections, we (1) discuss how our findings relate to prior work on the design of nonverbal cues in virtual environments, (2) introduce cue banks, an approach to address the challenge of overwhelming the users with too many types of nonverbal cues, and (3) discuss considerations for future AI-powered recognition of nonverbal cues.

\textbf{Another Step Towards VR Accessibility.} Prior work has explored nonverbal behaviors that are commonly used in social VR, such as head movements \cite{maloney2020talking, aburumman2022nonverbal} and hand gestures \cite{maloney2020talking, tanenbaum2020make}. Prior work has also emphasized the importance of nonverbal behaviors in interactions between users, such as that nodding between users and avatars resulted in higher levels of trust towards avatars \cite{aburumman2022nonverbal}. Our designs of accessible nonverbal cues add to this body of work to allow BLV users to receive information about nonverbal behaviors that are commonly used in a virtual environment. 

In prior work, Wieland et al. \cite{wieland2023vr} explored different modalities to represent eye gaze in VR and found that participants had positive responses to using audio and haptic to represent eye gaze. Our evaluation adds to Wieland et al.’s findings, showing that audio and haptic designs can be effective ways of representing nonverbal behaviors, but also that there is great variability in how people respond to these modalities in practice. For instance, nearly all participants could easily perceive haptics without being distracted from conversation. However, continuous vibrations can be overwhelming or overstimulating, even if they are easy to maintain in the background of a conversation. 

In another study, Wieland et al. \cite{wieland2022towards} identified which nonverbal behaviors are important to incorporate into conversations in VR, and suggested that multiple kinds of cues should be incorporated into conversation. While our work echoed this suggestion, and our participants suggested a variety of new cues to incorporate into future systems, we also found that including too many cues at once becomes quickly overwhelming in practice. Therefore, new methods of handling large sets of cues should be considered.

These suggestions are centered on making social VR accessible by supporting communication for BLV users. However, there are many other directions for future work to address social VR inaccessibility. For instance, while cue systems can help solve communication challenges in one-on-one conversations or small groups, future work can consider how these systems could support larger social venues in VR, such as a music concert \cite{vrchat-music-revolution}. Researchers should consider whether cue systems could provide information about large and complex groups without detracting from a BLV user’s enjoyment and immersion in these scenarios. 

\textbf{Cue Banks.} Both our participants and prior work suggested several cues to add to future systems. However, if all of these cues were implemented and provided to users at once, they would likely become overwhelming. Future work could explore options for handling large sets of cues in ways that will not overwhelm BLV users.

One way to address this would be “cue banks.” These can contain all possible nonverbal cues that the system can represent. Within these larger banks, there can be sub-banks of cues that are commonly used in different situations. Users could select and activate specific sub-banks to receive smaller sets of cues that they find relevant to their social situations. We have developed some example sub-banks which include cues that our participants suggested (see section \ref{Design Suggestions}), grouped with other possible cues we believe fall under similar categories of use:

\begin{itemize}
\item{Boredom cues: looking at a mobile device, glazed-over eyes, nodding off}
\item{Greeting cues: waving, holding out hand for handshake}
\item{Listening cues: eye gaze, head nodding, head shaking}
\item{Reaction cues: scowl, wide eyes of surprise, big smile}
\item{Subtle eye cues: judgemental stare, suspicious stare, interested stare}
\item{Facilitating cues: raising hands, confused expression, “boredom cues”}
\end{itemize}

As seen above, sub-banks would vary in size, such as the “facilitating cues” sub-bank which may contain smaller sub-banks like “boredom cues.” Future work could investigate the ideal sizes of sub-banks, or additional kinds of sub-banks that BLV users may find useful. Another approach the research community could examine is allowing users to apply sub-banks to specific people in specific social scenarios. For example, if a BLV user is giving a presentation to their coworkers and boss, they may be looking for different social information among the audience members. They might assign the “reaction cues” to their boss, so they know if their boss likes the ideas they are presenting. For everyone else, they might only attach the basic “listening cues” to ensure they are paying attention.

It is important to also consider how these sub-banks might be integrated with existing commercial VR platforms, such as VRChat and RecRoom. These existing platforms often include chaotic, uncontrolled environments, where such sub-banks may introduce challenges such as information overload due to the large number of users within the environment. 

Finally, rather than relying on the users themselves to assign and activate different sub-banks, future cue systems could attempt to “read the room,” and activate certain sub-banks based on the room’s liveliness or atmosphere. For example, if the room becomes unusually silent for an extended period of time, the cue system can automatically turn on “boredom cues” or “listening cues” to help the user assess the situation. If the room becomes lively again, the system could turn off the previous sub-bank and apply “reaction cues” instead. Supposing a new speaker approached the BLV user now that the mood lifted, the system could quickly turn on “greeting cues,” ensuring the user notices when the new speaker holds out a hand to introduce themselves.

While the above system designs have great potential, it is important to first acknowledge that there are cultural differences surrounding nonverbal cues. These differences may make the process of assigning sets of cues with specific meanings (such as “boredom” or “greeting” cues) much less certain. This study was conducted in the United States, where nonverbal cues may have very different meanings compared to how they are used in other countries. For example, while eye contact in the United States is commonly understood to represent someone paying attention, in other cultures such as some “Asian, African, and Latin American countries, eye contact should be avoided to show respect” \cite{nonverbal-communication-cultural}. Further, there may be cultural differences in social cues used by sighted and BLV people. While our work gave BLV people access to sighted-normative cues to support conversations, in a situation where a sighted person was in a room full of blind people, there might be BLV-normative cues of an entirely different kind that the sighted person would need to carry conversation. Future work should take cultural differences into account when designing and evaluating nonverbal cues in different countries or mixed-ability spaces. 

\textbf{Challenges with AI-Powered Detection of Nonverbal Behaviors.} Another important consideration for future cue systems is implementing AI in accessible VR, as the use of AI in VR is growing rapidly. Many of the challenges of implementing AI in VR are well-documented \cite{luck2000applying, cipresso2015virtual, ribeiro2023virtual}, so we focus our discussion here on the unique difficulties and opportunities we foresee with using AI for nonverbal behaviors. First, AI-based detection introduces challenges due to its inaccuracy with identifying many facial expressions and gestures. Current face recognition technologies are known to have difficulties determining even basic emotional expressions such as happiness or sadness with high accuracy, let alone subtle differences between types of eye gaze \cite{begel2020lessons}. One potential way to address this uncertainty may be to modify the nonverbal cue feedback. For example, when a user receives audio feedback that a conversational partner smiled, if the AI has 99\% certainty the event happened, the audio feedback could be higher in volume. In contrast, if the AI has a lower level of certainty, the audio feedback could be lower in volume. Haptic versions of these feedback systems could be applied as well, so that participants can receive confidence information via one channel (tactile feedback from haptics) while they receive the accessible cue itself via another (audio feedback from the accessible cue). This would allow users to quickly assess the certainty of the detection system as they converse.

In addition, there are possible privacy concerns with AI constantly tracking people’s nonverbal behaviors in order to implement such a system. Some users may feel uncomfortable with AI collecting such information, particularly if they are not made aware that the system is observing and translating their behavior for another user. One possible solution would be for other users to provide “consent” to share their nonverbal behaviors when a system tries to access them, although this could inadvertently expose the BLV user’s visual condition by telling users there is a person with a visual impairment nearby. Another possibility is to have the system only detect the behaviors of those on the BLV user’s friend list, or make consent to share nonverbal behaviors a part of the friend request system. In general, future work should take privacy concerns into account when creating systems that track nonverbal behaviors.
\section{Conclusion}
In this paper, we designed and evaluated accessible audio and haptic cue representations of three nonverbal behaviors: eye contact, head shaking, and head nodding. Our evaluation of the accessible nonverbal cues with 16 blind and low vision participants highlighted that the designs of the accessible nonverbal cues were effective in supporting conversations. We also uncovered a range of participant preferences for current cue designs and various approaches the research community could examine for future cue designs. Our work points to a novel avenue of making social VR fully accessible for blind and low vision people. 

% \input{sections/final-diff}

\bibliographystyle{ACM-Reference-Format}
\bibliography{references}

\end{document}